\documentclass[sigconf, manuscript]{acmart}

\usepackage{siunitx}
\usepackage{booktabs}
\usepackage{tabularx}
\usepackage{multirow}
\usepackage{calc}
\usepackage[capitalize, noabbrev]{cleveref}

\newcommand{\changenote}[1]{} %

\newcommand{\lastaccessed}{\textit{last accessed 04.01.2021}}

\newcommand{\pct}[1]{\SI{#1}{\%}}
\newcommand{\ms}[1]{\SI{#1}{ms}}
\newcommand{\secs}[1]{\SI{#1}{s}}

\newcommand{\langnonnative}{\overline{en}}

\newcommand{\suggnone}{s_0}
\newcommand{\suggone}{s_1}

\newcommand{\ttestcohend}[3]{t=#1, p#2, d=#3}

\newcommand{\ivlanguage}{\textit{language proficiency}}
\newcommand{\ivsuggestions}{\textit{parallel suggestions}}

\newcommand{\langnative}{en}
\newcommand{\suggthree}{s_3}
\newcommand{\suggsix}{s_6}

\newcommand{\hmmstate}[1]{St\textsubscript{#1}}

\newcommand{\glmmci}[5]{$\beta$=#1, SE=#2, CI$_{95\%}$=[#3, #4], p#5}

\definecolor{rowcol}{rgb}{0.9,0.9,0.9}

\makeatletter
\newcommand\footnoteref[1]{\protected@xdef\@thefnmark{\ref{#1}}\@footnotemark}
\makeatother

\AtBeginDocument{%
  \providecommand\BibTeX{{%
    \normalfont B\kern-0.5em{\scshape i\kern-0.25em b}\kern-0.8em\TeX}}}

\copyrightyear{2021}
\acmYear{2021}
\setcopyright{acmlicensed}\acmConference[CHI '21]{CHI Conference on Human Factors in Computing Systems}{May 8--13, 2021}{Yokohama, Japan}
\acmBooktitle{CHI Conference on Human Factors in Computing Systems (CHI '21), May 8--13, 2021, Yokohama, Japan}
\acmPrice{15.00}
\acmDOI{10.1145/3411764.3445372}
\acmISBN{978-1-4503-8096-6/21/05}

\begin{document}

\title{The Impact of Multiple Parallel Phrase Suggestions on Email Input and Composition Behaviour of Native and Non-Native English Writers}
\renewcommand{\shorttitle}{The Impact of Multiple Parallel Phrase Suggestions}

\author{Daniel Buschek}
\orcid{0000-0002-0013-715X}
\email{daniel.buschek@uni-bayreuth.de}
\affiliation{%
  \institution{Research Group HCI + AI, Department of Computer Science, University of Bayreuth}
  \streetaddress{Universit\"atsstra{\ss}e 30}
  \city{Bayreuth}
  \state{Germany}
  \postcode{95447}
}

\author{Martin Zürn}
\email{martin.zuern@campus.lmu.de}
\affiliation{%
  \institution{LMU Munich}
  \streetaddress{Frauenlobstra{\ss}e 7a}
  \city{Munich}
  \state{Germany}
  \postcode{80337}
}

\author{Malin Eiband}
\email{malin.eiband@ifi.lmu.de}
\affiliation{%
  \institution{LMU Munich}
  \streetaddress{Frauenlobstra{\ss}e 7a}
  \city{Munich}
  \state{Germany}
  \postcode{80337}
}

\renewcommand{\shortauthors}{Buschek et al.}

\begin{abstract}
  
We present an in-depth analysis of the impact of multi-word suggestion choices from a neural language model on user behaviour regarding input and text composition in email writing. 
Our study for the first time compares different numbers of parallel suggestions, and use by native and non-native English writers, to explore a trade-off of ``efficiency vs ideation'', emerging from recent literature.
We built a text editor prototype with a neural language model (GPT-2), refined in a prestudy with 30 people. In an online study (N=156), people composed emails in four conditions (0/1/3/6 parallel suggestions). 
Our results reveal (1) benefits for ideation, and costs for efficiency, when suggesting multiple phrases; (2) that non-native speakers benefit more from more suggestions; and (3) further insights into behaviour patterns.
We discuss implications for research, the design of interactive suggestion systems,  and the vision of supporting writers with AI instead of replacing them.
\end{abstract}



\maketitle

\section{Introduction}

More and more end-user applications use language modelling for real-time text predictions to support users interactively. In general, language models predict likely next words based on previous text. This is used, for example, in modern mobile on-screen keyboards (e.g. \textit{SwiftKey}\footnote{\label{footnote:swiftkey}\url{https://www.microsoft.com/en-us/swiftkey}, \lastaccessed}) with features such as word suggestions and autocorrection. 
Increasingly, such features leverage recent neural network-based language models from Natural Language Processing (NLP): For instance, Google's \textit{Smart Reply}~\cite{Kannan2016} and \textit{Smart Compose}~\cite{Chen2019} use such models to predict complete replies and phrases to reduce typing efforts in emails. 
More widespread use of such models in interactive systems and products is to be expected, as indicated by libraries for developers~\cite{Wolf2019huggingfaces} and system-as-a-service solutions\footnote{\url{https://inferkit.com}, \lastaccessed}.

\begin{figure}[!t]
\centering
\includegraphics[width=\minof{\columnwidth}{0.5\textwidth}]{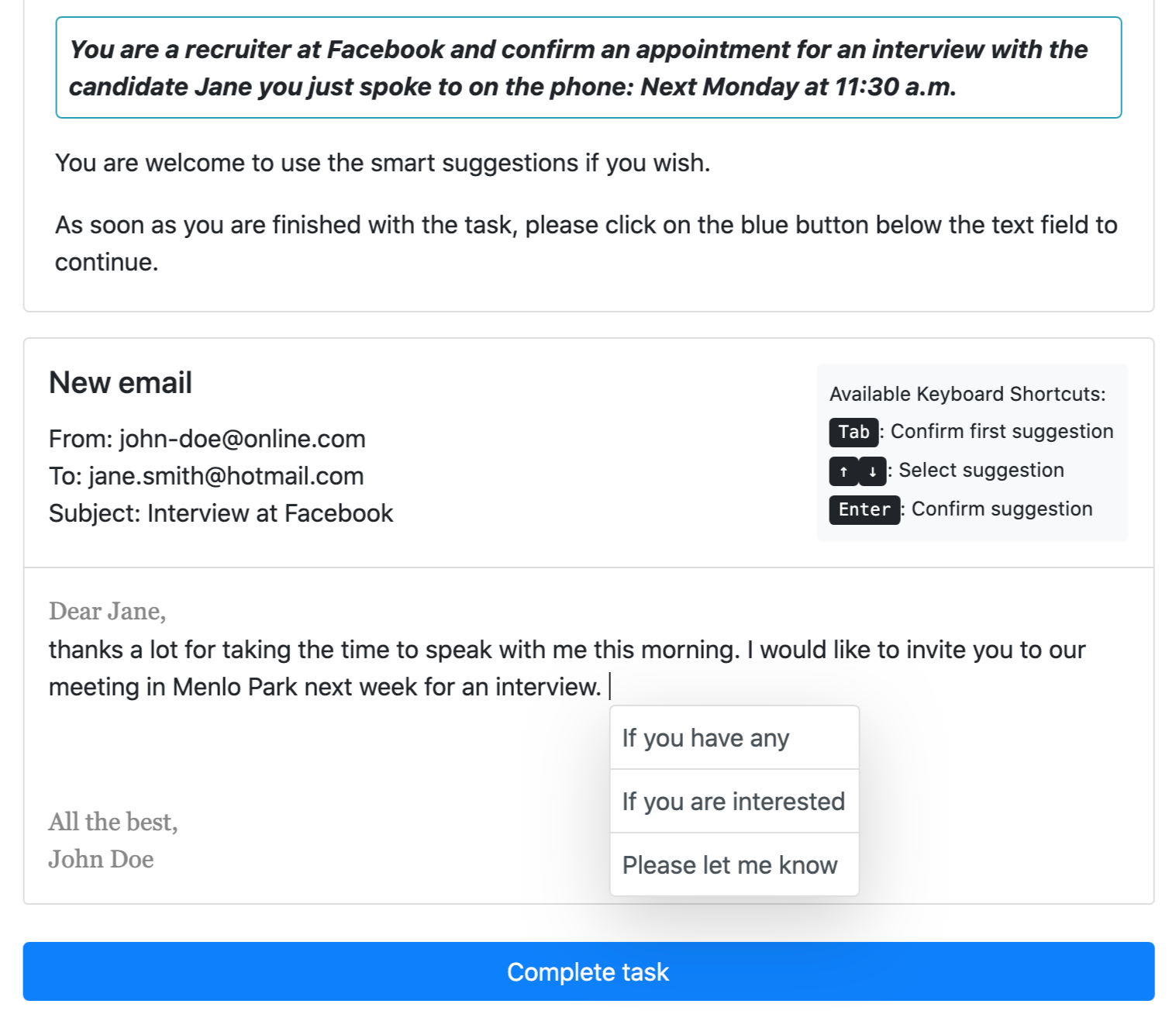}
\caption{Our task UI and prototype email web app with multiple parallel multi-word (i.e. ``phrase'') suggestions generated by a neural language model (\textit{GPT-2}). Interaction with the suggestion list is possible via mouse or keyboard (tab to confirm first suggestion, arrow up/down to change selection, enter to accept selected suggestion).}
\label{fig:teaser}
\Description[Screenshot of text editor with email and suggestions.]{Screenshot of text editor showing partly entered email and list of three suggestions as popup at the cursor location.}
\end{figure}

Ongoing NLP research improves and scales up such models for text generation (e.g.~\cite{Brown2020language}), yet complementary investigations from a Human-Computer Interaction (HCI) perspective remain sparse, that is, regarding \textit{user interaction} with systems using such models and their impact on user behaviour and resulting text.

Text prediction in HCI often aims to improve typing speed and reduce errors~\cite{Kristensson2014} or keystrokes~\cite{Chen2019}. However, work by \citet{Arnold2016uist}, for example, suggests that there is more to text predictions than efficiency: 
People interpreted multi-word phrases as suggestions for content and expression, while single words were seen as predictions of typing.  
\citet{Arnold2016uist} here referred to impacts on ``process'' and ``product'', which points to an emerging trade-off %
for design: %
On the one hand, short suggestions may be designed for \textit{efficiency}, to reduce keystrokes and typing time, yet provide no composition ideas. On the other hand, reading more and longer suggestions takes time yet may help users with coming up with what to write next and how to write it (which we refer to as \textit{ideation} for short). %
To investigate this, we explore this research question: 
\changenote{shortened} %
\textit{How do different numbers of parallel suggestions affect input and text composition behaviour of users with native and non-native language proficiency?}

In addressing this, we cover two new vital aspects: 
(1)~On the system side, the \textit{number of parallel suggestions} shown is a key design factor, as indicated by prior work~\cite{Arnold2016uist, Nicolau2019, Quinn2016chi}, without empirical study so far. %
(2)~On the user side, desired design choices %
might be informed by skill, such as typing and \textit{language proficiency}, an important factor in writing proficiency~\cite{Leijten2019}. To our knowledge, language proficiency has not been studied yet in the context of phrase suggestions.

We built a text editor prototype (\cref{fig:teaser}) with the language model \textit{GPT-2}~\cite{Radford2019}\footnote{\label{footnote:gpt2}\url{https://huggingface.co/gpt2}, \lastaccessed}, refined in a prestudy (N=30). We then logged user interactions in an email writing study online (N=156). %
As key results, our analyses reveal the benefits and costs of suggesting multiple phrases, differences in suggestion use between native and non-native speakers, and further insights into behaviour patterns.

We release the model, data, and study material to facilitate further research.
In this way, we hope to support the vision of augmenting human writers with AI, instead of replacing them.

\section{Related Work}

We relate our work to research from NLP and HCI. %

\subsection{Interpreting and Correcting Typing Input}\label{sec:intepreting_input}

A longstanding use of language models in text entry is input interpretation to improve speed and reduce errors in mobile keyboards~\cite{Kristensson2014}:
Early work disambiguated input with word frequencies (e.g. T9)~\cite{James2000}. Later, \citet{Goodman2002} combined a seven-gram character language model with stylus taps on a PDA. %
Modern smartphone keyboards (e.g. \textit{SwiftKey}\footnoteref{footnote:swiftkey}) and related work (e.g.~\cite{Goel2012, Vertanen2015, Yin2013}) use this combination to infer text from touches. %
Language information can also help to correct words \textit{after} typing (``autocorrect''~\cite{Banovic2019mobilehci}).
In contrast, we focus on predicting phrases, in desktop typing, generated with a neural model. %
While n-gram predictions only consider the last n-1 words, neural models can consider larger context.

\subsection{Predicting Text Input}\label{sec:predicting_input}

Language models are also used to reduce typing: For instance, many smartphone keyboards suggest words, including both completions (e.g. ``conference'' after typing ''conf'')~\cite{Bi2014} and next words.

Many systems suggest \textit{one} next word (e.g.~\cite{Dunlop2012, Fowler2015, Gordon2016, Quinn2016chi}). Likely motivations for this design choice include limitations in (n-gram) models, screen space and interactions~\cite{Gordon2016}, and cost-benefit trade-offs for efficiency (e.g. reading time vs saved keystrokes)~\cite{Palin2019, Quinn2016chi}.

However, recent work has explored designs beyond one word: 
For example, both Google's \textit{Smart Compose}~\cite{Chen2019} and \textit{Smart Reply}~\cite{Kannan2016} suggest phrases. \textit{Smart Compose}~\cite{Chen2019} shows the most likely sentence completion if a confidence threshold is passed. \textit{Smart Reply}~\cite{Kannan2016} shows an optimised set of two to three options (e.g. ensuring to include a positive and negative reply suggestion). %

A key motivation for our work is to investigate interaction with systems which use neural language models to go beyond single words. In particular, we analyse the impact on input and text composition behaviour. %
Predictive features may influence composition, which motivates us to compare native and non-native language writers, for whom cognitive processes and efforts differ (e.g. see~\cite{Leijten2019}).

\subsection{Perception and Impact of Suggestions}

There is more to predictive text entry than saving keystrokes: \citet{Arnold2016uist} compared suggesting phrases vs single words on a mobile keyboard in a restaurant review task. People interpreted phrases as suggestions on content or how to express something. In contrast, single words were seen more as predictions. 

Such insights motivate our work in that they hint at an emerging trade-off of efficiency vs inspiration through predictive text systems. %
In follow-up work, \citeauthor{Arnold2017counterfactual} focused on technical improvements for phrase suggestions~\cite{Arnold2017counterfactual}, and on bias, for example, in image descriptions~\cite{Arnold2020} and review sentiment~\cite{Arnold2018}. 

In contrast, we focus on how such suggestions influence user behaviour (e.g. suggestion acceptance and modification, sequential behaviour in text composition, user perception). %
Moreover, instead of mobile use, we study email writing at the desktop, where typing tends to be faster already so that the inspirational use of suggestions, rather than efficiency, might be more practically interesting. %

\subsection{Generating Text and Interactive NLP}

In NLP, new Deep Learning architectures~\cite{Vaswani2017nips}, pre-training~\cite{Qiu2020pretrained}, and growing datasets, models, and computing power (e.g.~\cite{Brown2020language}) have improved many benchmarks, including language models for understanding and generating text~\cite{Brown2020language, Devlin2019bert}. 
Some of this work inspired popular interactive demos, yet without empirical analysis of their use and UI design (e.g.\textit{Write With Transformer}\footnote{\label{footnote:write_with_transformer}\url{https://transformer.huggingface.co/}, \lastaccessed}). 

Further (HCI) work on interactive text generation looked into artistic use, with roles akin to creativity-support~\cite{Frich2019}, such as for poetry~\cite{Hafez2017}, metaphors~\cite{Gero2019chi}, slogans~\cite{Clark2018}, and stories~\cite{Tambwekar2018controllable}. 
Beyond artistic use, a 2019 survey on academic writing tools found that only nine out of 44 tools use NLP~\cite{Strobl2019}, motivating further work. %

In this interdisciplinary context, we contribute a detailed HCI analysis for neural language models as suggestion providers regarding their role and impact on user behaviour in text entry systems. %
In a wider view, we thus situate our work among growing research interests in using NLP in HCI~\cite{Yang2019}, such as for summarisation~\cite{Gehrmann2020}, document understanding~\cite{ter_Hoeve2020}, and conversational UIs~\cite{Candello2020}.

In a visionary perspective, understanding user interaction with neural language models is motivated by the vision of augmenting what human writers can do, instead of replacing them. Supporting this, %
a report on the social impact of recent language models tied beneficial use also to future improvements in \textit{user interfaces}~\cite{Solaiman2019release}. %

\subsection{UI Design Factors for Text Suggestions}\label{sec:ui_design_factors}

Explicit enumerations of UI design factors for text suggestions are sparse: \citet{Nicolau2019} presented a design space with seven dimensions for non-visual presentation of word completions, and \citet{Quinn2016chi} listed four dimensions for visual word suggestions. These included, for example, confidence representation~\cite{Nicolau2019} and the screen location~\cite{Quinn2016chi}. In their intersection, both listings included the number of suggestions displayed at a time.

Numbers of suggestions seem to have been chosen based on UI space (e.g. five on desktop~\cite{Trnka2009}, three on smartphones~\cite{Nicolau2019}, two on watches~\cite{Gordon2016}). Even simulation experiments for smartphone keyboards fixed the number to three~\cite{Fowler2015}. Considering products, \textit{Smart Reply} in Gmail also shows two to three suggestions~\cite{Kannan2016}.
At the same time, related work highlighted the number of suggestions as an important design factor without empirical study~\cite{Arnold2016uist, Quinn2016chi}.

This gap motivates our detailed analysis of this factor here. In addition, we deem the number of suggestions particularly relevant for investigating the described trade-off between efficiency and ideation: Reading more suggestions takes time yet may also include more potentially inspiring content. To the best of our knowledge, this is the first reported study of a phrase suggestion system to investigate this fundamental UI design factor empirically.

\subsection{Writing Research}

Cross-references between writing research and HCI work on text entry (cf. \cref{sec:intepreting_input}, \ref{sec:predicting_input}) are sparse, despite overlapping interests, not least in methodology: Interaction logs are not only a key instrument in HCI but also in writing research~\cite{Leijten2013, Lindgren2019}.
Insights into cognitive processes in writing~\cite{Deane2008, Flower1981} seem particularly relevant for studying systems, such as text generators, that may influence how writers come up with ideas, express them, and revise them~\cite{Mahlow2014}. 
As one example, cognitive efforts and processes differ between writing in a native or foreign language, which shows down to the keystroke level~\cite{Leijten2019}. 
This motivates us to study the impact of (different numbers of) suggestions for both native and non-native English writers and also to examine interaction logs in detail.

\section{Prototype Text Suggestion System}
\label{sec:prototype}
We developed a web app with an email composition UI (Figure~\ref{fig:teaser}). %

\subsection{User Interface and Interactions}\label{sec:prototype_ui}

The frontend of the prototype was implemented using React\footnote{\url{https://reactjs.org/}, \lastaccessed}.

\subsubsection{Main GUI}
The main study UI (\cref{fig:teaser}) shows the scenario and email field. Email header/greetings were pre-filled to save time; the signature used an anonymous nickname that the app asked for beforehand. The signature did not limit the text (textbox grows when adding lines). The blue button was used to submit the email. 

\subsubsection{Interactions}
The email field supported all typical interactions, such as typing, selecting, editing, and deleting text, and caret movement via keys and mouse. Selecting suggestions in the list was possible via \textit{mouse (point and click)} or \textit{keyboard (tab to confirm first suggestion, arrow up/down to change selection, enter to accept selected suggestion)}. These key commands were explained as part of the study and shown in the UI (see \cref{fig:teaser}).

\subsubsection{Suggestions}
We call a suggestion ``accepted'' if the user selected it from the list to add it to the text. A list is ``dismissed'' if the user accepted none of its suggestions. For example, a list can be dismissed by continuing to type.
The list pops up at the caret (\cref{fig:teaser}), \ms{250} after the last keypress. %
Recent work argued against delays in marking menus~\cite{Henderson2020}, yet we used a small one here to reduce `` UI flicker'' and computational load, for a more consistent experience, since ``instant'' updates were not possible in our prototype due to the computational costs.
Visually, we chose a popup look and placement below the current line for clear visual distinction of suggestions and the user's text. 
Overall, our UI/interaction design (i.e. pop-up look/placement, selection interactions, loading indicator) was informed by related designs (cf. \textit{Write With Transformer}\footnoteref{footnote:write_with_transformer}; suggestions in IDEs, e.g. \textit{TabNine}\footnote{\url{https://www.tabnine.com/}, \lastaccessed}).
We do not claim this to be ``optimal'' and reflect on further ideas in the discussion.

\subsection{Backend and Deployment}

Our backend is written in Python and is used by the frontend both to request text suggestions and to log data. %
In the prestudy, the backend integrated the model directly, which we changed for the main study to improve performance (see Section~\ref{sec:prototype_after_prestudy}). %
The logs were written to an ElasticSearch cluster. %
We deployed the system on AWS infrastructure on an Elastic Container Service (ECS) cluster.%

\subsection{Language Model}

We used the pre-trained \textit{GPT-2}~\cite{Radford2019} for English via HuggingFace %
(see model cards\footnoteref{footnote:gpt2}). %
We finetuned it for our email use case on ENRON~\cite{Klimt:2004} %
(the lightly preprocessed version by Brian Ray\footnote{\url{https://data.world/brianray/enron-email-dataset}, \lastaccessed}).
We removed very short messages (e.g. informal one-liners), automated system messages, quoted replies and signatures, leaving 236,206 emails. We replaced dates and names with placeholders (which the backend filled in based on the study scenario). %
For finetuning, we used HuggingFace's \textit{GPT-2} training script. %
Suggestions are generated via beam search\footnote{e.g. see \url{https://huggingface.co/blog/how-to-generate}, \lastaccessed} (twice as many beams as suggestions and 1.5 repetition penalty) and shown in order of likelihood.
Thus, we show ``top'' suggestions, not an optimised \textit{set} (see~\cite{Deb2019, Kannan2016}). %

\section{Prestudy}\label{sec:prestudy}

To iterate on our prototype, we tested it in an online prestudy.%
\subsection{Study Design and Apparatus}\label{sec:prestudy_design_apparatus}

The study was approved by our institution. %
We used our\textit{ web app} (\cref{sec:prototype}) and logged a broad range of data, including suggestion loading and selection time, viewed, accepted and rejected suggestions, and more. %
We designed three email tasks (\textit{scenarios}) with typical business emailing in mind, yet leaving room for people to include their own ideas (\cref{tab:email_scenarios}). %

A \textit{task questionnaire} after each task assessed preferred selection (mouse, tab, arrows + enter) and number/length of suggestions, and perception (Likert items -- Q1: \textit{``I feel that I am the author of the email.''}, Q2: \textit{``Due to the suggestions, I used phrases and words that I would not have used on my own.''}, and Q3: \textit{``I would use suggestions as shown here for daily use.''}).
A \textit{final questionnaire} assessed demographics, prior experience, and open feedback.

\subsection{Participants and Procedure}
We recruited 30 people (15 female, 15 male) via social media and word-of-mouth with a mean age of 26 years (range 22 - 33). %
25 had a Bachelor's degree or higher, 23 had an English level of C1\footnote{\label{footnote:cefr_scale}CEFR scale: \url{https://www.coe.int/en/web/common-european-framework-reference-languages/table-1-cefr-3.3-common-reference-levels-global-scale}, \lastaccessed}  or higher.  
People used their own laptop or desktop computer. They were informed about the study goal, provided consent, and chose a nickname for the email signature (see \cref{sec:prototype_ui}), before using our prototype. The study had two parts and took 10 to 15 minutes.

\subsubsection{Part 1: Fixed Settings}\label{sec:prestudy_part1} 
People were randomly assigned to one of our three scenarios (\cref{tab:email_scenarios}). We set the number of suggestions to four and words to up to four (i.e. <4 if the model predicts end-of-sentence before reaching 4). These values were informed by the design of many current systems (see related work, Section~\ref{sec:ui_design_factors}). %
Upon task completion, participants filled in the task questionnaire.

\subsubsection{Part 2: Customisable Settings}\label{sec:prestudy_part2} 
People were randomly assigned another (different) one of the three scenarios. Now, we encouraged them to freely explore any suggestion numbers (up to 20) and words per suggestion (up to 15). The UI showed these settings between task description and the email field such that it was always accessible. People then filled in both the task and final questionnaire.

\begin{table}[t]
\centering
\footnotesize
\renewcommand{\arraystretch}{1.75}
\setlength{\tabcolsep}{4pt}
\begin{tabularx}{\linewidth}{@{}lXl@{}}
\toprule
\textbf{Scenario} & \textbf{Setting \& Task} & \textbf{Study} \\ 
\midrule
Birthday & You are a direct report to Anna, who is on a business trip abroad. Today is her birthday, so congratulate her via email and add some personal wishes. & pre, main \\
Reference & You need to write an email to your former peer Jane, asking her if she would be available as a reference for an application at Google. She recently got promoted to Manager, which you should mention. & pre, main \\
ID Card & Your ID card is expiring soon, so you need to write an email to your local city administration to get a new one. Please explain why you need to come by train and ask for directions. & pre, main \\
Interview & You are a recruiter at Facebook and confirm an appointment for an interview with the candidate Jane you just spoke to on the phone: Next Monday at 11:30\,am. & main \\ 
\bottomrule
\end{tabularx}
\Description[Table listing the study prompts.]{Table listing the study prompt texts (i.e. scenario texts) used in the prestudy and main study, as indicated by the last column.}
\caption{Email scenarios in the prestudy and main study.}
\label{tab:email_scenarios}
\end{table}

\subsection{Results and Discussion}

\subsubsection{Logging}
Emails contained a mean of 49.22 words (SD 17.35) with 14.53 (SD 12.11) from suggestions. People triggered 3042 suggestion lists with a mean loading time of \ms{1142} (SD \ms{711}). They accepted 455 suggestions with a mean selection time of \ms{4344} (SD \ms{4481}).
In the second part of the study, people mainly explored the maxima and minima for the number and length of suggestions. %

\subsubsection{Subjective Feedback}
People's preferred \textit{number of suggestions} had a median of 3 (M 3.67, SD 0.96) after the study's first part, and a median of 4 (M 3.9, SD 1.06) after the second.
The preferred \textit{number of words} had a median of 3 (M 3.0, SD  1.72) after the first part, and also a median of 3 (M 3.33, SD 1.42) after the second.

For the Likert items (see \cref{sec:prestudy_design_apparatus}), for the first part, people rated a median of ``agree''/4 for Q1 (feeling to be the author), ``agree''/4 for Q2 (using other phrases/words due to suggestions), and ``neutral'' to ``agree''/3.5 for Q3 (using such suggestions daily).

In the open feedback, 12 out of 30 people highlighted that the suggestions provided ideas on what to write; nine said they liked the UI; eight praised the suggestion quality; three mentioned help with text correction; one mentioned saving time. %
Top criticisms were long loading times of suggestions (13), insufficient suggestion quality (11), and lack of relevant suggestions for email writing (6).

\subsubsection{Summary and Learnings}

The results were encouraging with regard to our interest in the potential role of neural text suggestions for \textit{ideation}, as the open comments of 12 of 30 people related to this aspect. %
Users took about four seconds to select suggestions, which also indicates that they used them not mainly with a concern of optimising efficiency. We also received positive feedback on usability and UI.
Moreover, perceived ``ideal'' values for number and length of suggestions fell close to our defaults (3-4 suggestions). Note that these might be influenced by what people know (e.g. from smartphone keyboards) and by the defaults (part 1). However, people indeed explored settings (part 2) and some changed their preference, indicating reflection on the experience and settings.

\subsection{Improvements After the Prestudy}\label{sec:prototype_after_prestudy}
We identified two main points for improvement:
First, suggestions took too long to load. %
Second, we received mixed feedback on suggestion quality: While many participants perceived them as convenient, others criticised their lack of precision (e.g. context relevance). Thus, we improved our prototype. %

For speed up, we moved the model to AWS SageMaker\footnote{\url{https://aws.amazon.com/sagemaker/}, \lastaccessed}.
We also added the email subject as context for the model (prepended with delimiter token, see \cite{Wolf2019}).
Finally, we improved logging details, fixed a bug with placeholder tokens, and found further emails to exclude (e.g. further log messages). We then repeated the finetuning. %

\section{Main Study}\label{sec:main_study}

\subsection{Study Design and Apparatus}

The study was approved by our institution. It had a within-subject design for the independent variable \ivsuggestions{} with four levels: no suggestions ($\suggnone$), and one, three, six suggestions ($\suggone$, $\suggthree$, $\suggsix$).  
We studied \ivlanguage{} between subjects, with two levels: $\langnative$ (native English) and $\langnonnative$ (non-native).
As dependent variables we logged a wide range of measures (cf. prestudy and \cref{fig:metrics_overview}). %

\subsubsection{Web App and Scenarios}

We used our improved prototype (\cref{sec:prototype_after_prestudy}). %
While median preferred words per suggestion in the prestudy was three, the mean was slightly higher after people tried other settings. We decided to use up to four words, also motivated by our focus on suggesting phrases.  
We reused the three prestudy scenarios and added a fourth one to match our within-subject design (4 levels), see \cref{tab:email_scenarios}. %
The system enforced that emails were at least ten words long, yet we required no minimum writing time.

\subsubsection{Questionnaires}
People filled in a \textit{task questionnaire} after each task (email). %
Beyond the prestudy questions, we added Likert items on more detailed perceived impact (\textit{``The text suggestions influenced the content of the email.''} and \textit{``The text suggestions influenced the wording of the email.''}).
The study concluded with a \textit{final questionnaire} as in the prestudy, also including open feedback.%

\subsection{Participants and Procedure}

We recruited 162 people via \textit{Prolific}\footnote{\url{https://www.prolific.co/}, \lastaccessed}. We excluded six whose data showed that they had composed the text externally without our system. Thus, analyses are based on the remaining 156 participants.

These people (66 female, 89 male, 1 prefer not to disclose) had a mean age of 28 years (range: 18 - 72). Most lived in the UK (\pct{39.1}), the US (\pct{13.5}), Poland (\pct{10.9}), Portugal (\pct{9.6}), and Italy (\pct{5.8}).
\pct{51.9} were employed, \pct{26.9} students, \pct{11.5} students working part-time, \pct{8.3} unemployed, and \pct{1.3} retired. 
Common occupations included IT, Education, Finance \& Business, Retail, Engineering, Healthcare, and Media \& Arts.

We varied \textit{Prolific's} setting for requesting English native speakers to get varied proficiency (CEFR scale: A1 \pct{2.6}, A2 \pct{3.8}, B1 \pct{6.4}, B2 \pct{14.1}, C1 \pct{19.2}, C2 \pct{13.5}, native \pct{40.4}). An explanation of the scale and a link\footnoteref{footnote:cefr_scale} was provided.

\textit{Prolific} asks for fixed pay (in \pounds). %
We offered \pounds\,3.13 based on their recommendation, estimated duration, and US minimum wage. With a median completion time of 21.5 minutes effective compensation was \pounds\,8.75 (\$\,11.19) per hour.

People completed four email tasks, with a questionnaire after each, plus a final one. %
Each task used one of the scenarios and one of the \ivsuggestions{} conditions (0/1/3/6). These were counterbalanced with two orthogonal $4 \times 4$ Latin squares~\cite{McKay2007} and thus covered all combinations of scenarios/suggestions, and all scenarios/suggestions occurred at all positions in the task order.

\begin{figure*}[!t]
\centering
\includegraphics[width=\textwidth]{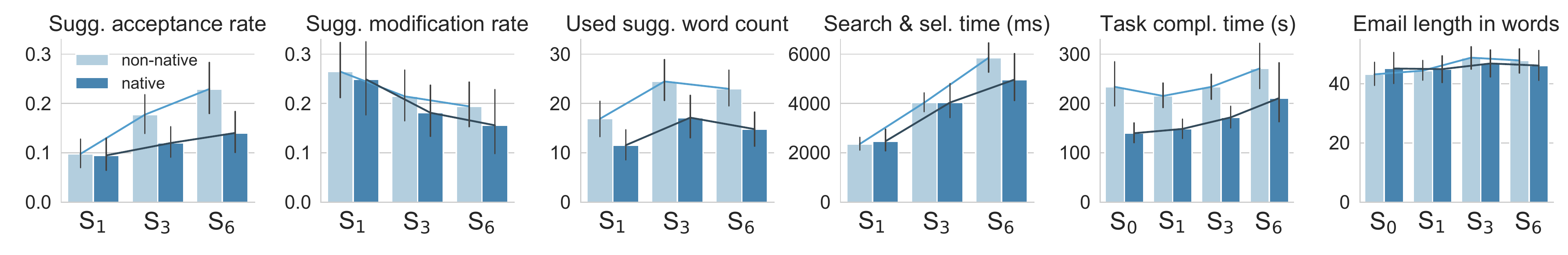}
\vspace{-2.5em}
\caption{Overview of key metrics. Bars show means (\pct{95} CIs), split by \ivlanguage (colour) and \ivsuggestions (x-axis). In case of multiple values per person per task (i.e. search \& selection times) values were averaged per person per task first such that each person contributes one value per level of \ivsuggestions.} %
\label{fig:metrics_overview}
\Description[Six bar charts showing measures from logged data.]{Six bar charts. Bars are split by native vs non-native, and by suggestion number. From left to right: suggestion acceptance rates with linear increasing trend for higher suggestion number, and non-natives higher than natives; suggestion modification rates with decreasing trend for higher suggestion number, with non-natives slightly higher; used suggestion word counts with S1 less than S6 less than S3, and non-natives higher than natives; search and selection times with linear increasing trend for higher suggestion number, and non-natives higher than natives; task completion times, which are roughly stable across suggestion numbers, yet higher for non-natives; and email length in words, also roughly stable and similar for natives and non-natives.}
\end{figure*}

\section{Results}

We used R~\cite{R2020} for significance testing, mainly (generalised) linear mixed-effects models (LMMs, packages \textit{lme4}~\cite{Bates2015} and \textit{lmerTest}~\cite{Kuznetsova2017}). The LMMs accounted for individual differences via random intercepts (for participant and scenario), plus the fixed effects (\ivlanguage{} and \ivsuggestions).
In addition, for Likert (ordinal) data, we used Generalized Estimating Equations (GEEs, R package \textit{multgee}~\cite{multgee2015}). We report significance at p<.05.

\subsection{Dataset Overview}
The emails contained 28,670 words with a mean of 45.95 words per email (SD 19.00), written in 3.50 minutes (SD 2.90).
The mean loading time per suggestion list was \ms{361} (SD \ms{243}) -- less than a third of the time in the prestudy, showing that our changes (\cref{sec:prototype_after_prestudy}) improved the prototype's speed drastically.
People accepted 3,920 suggestions. In tasks with suggestions, they accepted a mean of 8.38 suggestions per email (SD 8.70, %
min 0, max 65). In 411 of 468 emails written in these tasks, people accepted at least one suggestion (\pct{87.82}). 
\cref{fig:metrics_overview} gives a descriptive overview. %

\subsection{Use of Suggested Text}\label{sec:use_of_suggested_text}
Here we analyse key metrics for the suggestion conditions ($\suggone, \suggthree, \suggsix$).%

\subsubsection{Suggestion Acceptance Rate}
The suggestion acceptance rate is the number of times a suggestion was accepted (and kept in the final text) divided by the number of times a suggestion list was shown to the user. The grand mean acceptance rate was \pct{14.80}. 
Descriptively (\cref{fig:metrics_overview}), non-native speakers had a higher acceptance rate than native speakers ($M_{\langnonnative}$=.17; $M_{\langnative}$=.12), and more parallel suggestions increased acceptance ($M_{\suggone}$=.10, $M_{\suggthree}$=.15, $M_{\suggsix}$=.19).

For significance testing, we fitted a generalised LMM on the binomial acceptance data (i.e. accepted and kept? yes/no, per shown list).
The model had $\suggthree$ and $\suggsix$ as significant positive predictors ($\suggthree$: \glmmci{.40}{.10}{.21}{.59}{<.0001}; $\suggsix$: \glmmci{.40}{.10}{.20}{.60}{<.0001}). Thus, showing more than one suggestion in the list significantly increased the chance of accepting one ($\suggthree$: $\exp(\beta)=1.43$ i.e. chance +\pct{49}; also $\suggsix$: +\pct{49}).
While \ivlanguage{} alone was not a significant predictor in this model (p=.74), interactions were significant and positive ($\suggthree$: \glmmci{.31}{.12}{.08}{.54}{<.01}; $\suggsix$: \glmmci{.65}{.12}{.41}{.89}{<.0001}): More parallel suggestions increased the chance of acceptance significantly more for non-native speakers than native ones ($\suggthree$: +\pct{36}; $\suggsix$: +\pct{92}).

\subsubsection{List Selection}

For $\suggthree$ and $\suggsix$, we further examined selection per shown suggestion with LMMs with list position and length in characters as predictors.
Position was significant and negative (all p<.0001): Suggestions shown lower down the list had a smaller chance of being selected.
For $\suggthree$, \pct{7.3} of suggestions shown at 1st place in a list were selected (2nd \pct{5.6}, 3rd \pct{4.0}); similarly for $\suggsix$ (1st \pct{5.8}, 2nd \pct{4.1}, 3rd \pct{2.8}, 4th \pct{2.2}, 5th \& 6th \pct{1.6}).
Suggestion length was also significant ($\suggthree$: \glmmci{.020}{.003}{.013}{.027}{<.0001}; $\suggsix$ \glmmci{.012}{.002}{.007}{.016}{<.0001}): A character increases selection chance by (rel.) \pct{2.0} ($\suggthree$) and \pct{1.3} ($\suggsix$).

\subsubsection{Suggestion Modification Rate}

The modification rate is the number of times an accepted suggestion was manually modified (e.g. user partly deletes it) divided by the number of times a suggestion was accepted by the user.
Descriptively (\cref{fig:metrics_overview}), more parallel suggestions had lower rates ($M_{\suggone}$=.26, $M_{\suggthree}$=.20, $M_{\suggsix}$=.18), with only small differences between native/non-native speakers ($M_{\langnative}$=.20, $M_{\langnonnative}$=.22).
We fitted a generalised LMM on the binomial modification data (i.e. modified? yes/no, per accepted suggestion).
It had $\suggsix$ as a significant negative predictor (\glmmci{-1.06}{.25}{-1.57}{-.57}{<.0001}): Showing six suggestions in the list significantly reduced the chance of modifying an accepted suggestion compared to one suggestion ($1-\exp(\beta)=.65$ i.e. chance -\pct{65}). While not significant for three (p=.05), the estimate (-\pct{34}) fits the picture that more suggestions lead to fewer modifications. 
Here, \ivlanguage{} was not a significant predictor (p=.88), and none of the interactions were significant ($\suggthree$: p=.76;  $\suggsix$: p=.09).

\subsubsection{Used Suggestions Word Count}
We analysed the number of words in the final email text inserted by suggestions:
Descriptively (\cref{fig:metrics_overview}), emails written with multiple parallel suggestions contained more suggested words ($M_{\suggone}$=14.74, $M_{\suggthree}$=21.50, $M_{\suggsix}$=19.67), and emails by native speakers contained fewer suggested words ($M_{\langnative}$=14.49, $M_{\langnonnative}$=21.44).

We fitted a generalised LMM (Poisson family) on the word count data.
The model had \ivlanguage{} as a significant predictor (\glmmci{.41}{.18}{.05}{.78}{<.05}): Thus, non-native proficiency, all else equal, was estimated to significantly increase an email's number of suggestion words by +\pct{51}. 
The model also had $\suggthree$ and $\suggsix$ as significant positive predictors ($\suggthree$: \glmmci{.41}{.05}{.32}{.50}{<.0001}; $\suggsix$: \glmmci{.24}{.05}{.14}{.34}{<.0001}). Thus, showing more than one suggestion in the list significantly increased the count of suggestion words in the email ($\suggthree$: +\pct{51}; $\suggsix$: +\pct{27}). The interactions were not significant ($\suggthree$: p=.68;  $\suggsix$: p=.19).

\subsubsection{Email Length}
We found no significant differences ($M_{\langnative}$=45.77, $M_{\langnonnative}$=46.07; $M_{\suggnone}$=43.97, $M_{\suggone}$=44.63, $M_{\suggthree}$=48.01, $M_{\suggsix}$=47.17).

\subsection{Task Completion and Selection Times}\label{sec:task_and_selection_times}

\subsubsection{Task Time}

We measured task time from starting a task to submitting the email (this includes reading the scenario). 
We fitted an LMM on this data. It had \ivlanguage{} as a significant positive predictor (\glmmci{95}{28}{41}{148}{<.001}): Native speakers were significantly faster ($M_{\langnative}$=\secs{168}, $M_{\langnonnative}$=\secs{238}).

In the model, \ivsuggestions{} was a positive predictor for all levels and significant for $\suggsix$ (\glmmci{73}{27}{21}{125}{<.01}). More parallel suggestions took more time, and writing with one suggestion was slightly faster than writing without suggestions ($M_{\suggnone}$=\secs{196}, $M_{\suggone}$=\secs{188}, $M_{\suggthree}$=\secs{209}, $M_{\suggsix}$=\secs{247}). However, these differences were only significant for the pairwise comparisons of $\suggnone$ vs $\suggsix$ (\ttestcohend{-3.092}{=.013}{-.248}) and $\suggone$ vs $\suggsix$ (\ttestcohend{-3.372}{=.005}{-.270}). 

Interactions were not significant. Descriptively, non-native speakers were faster with one suggestion than without suggestions (\secs{215} vs \secs{234}), while native speakers on average were slightly slower with one suggestion than without (\secs{148} vs \secs{140}).

\subsubsection{Search \& Selection Time}
We measured the time for suggestion viewing, search \& selection between showing a list and accepting a suggestion.
We fitted an LMM and found no significant effect of \ivlanguage{}  ($M_{\langnative}$=\ms{3673}, $M_{\langnonnative}$=\ms{4089}) and no significant interaction. Descriptively, non-native speakers took longer to choose from six suggestions ($M_{\langnative}$=\ms{4818} vs $M_{\langnonnative}$=\ms{5947}).

The model had \ivsuggestions{} as significant positive predictors ($\suggthree$: \glmmci{1543}{396}{771}{2315}{<.0001}; $\suggsix$: \glmmci{2510}{389}{1751}{3269}{<.0001}): As expected, longer suggestion lists resulted in longer search \& selection times ($M_{\suggone}$=\ms{2403}, $M_{\suggthree}$=\ms{4034}, $M_{\suggsix}$=\ms{5501}).

\subsection{Sequential Input Patterns}

\newcommand{\actionArrow}{arrows} %
\newcommand{\actionBackspace}{delete} %
\newcommand{\actionDismiss}{dismiss}
\newcommand{\actionKey}{key}
\newcommand{\actionSelect}{accept} %

\begin{figure*}[!t]
\centering
\includegraphics[width=\textwidth]{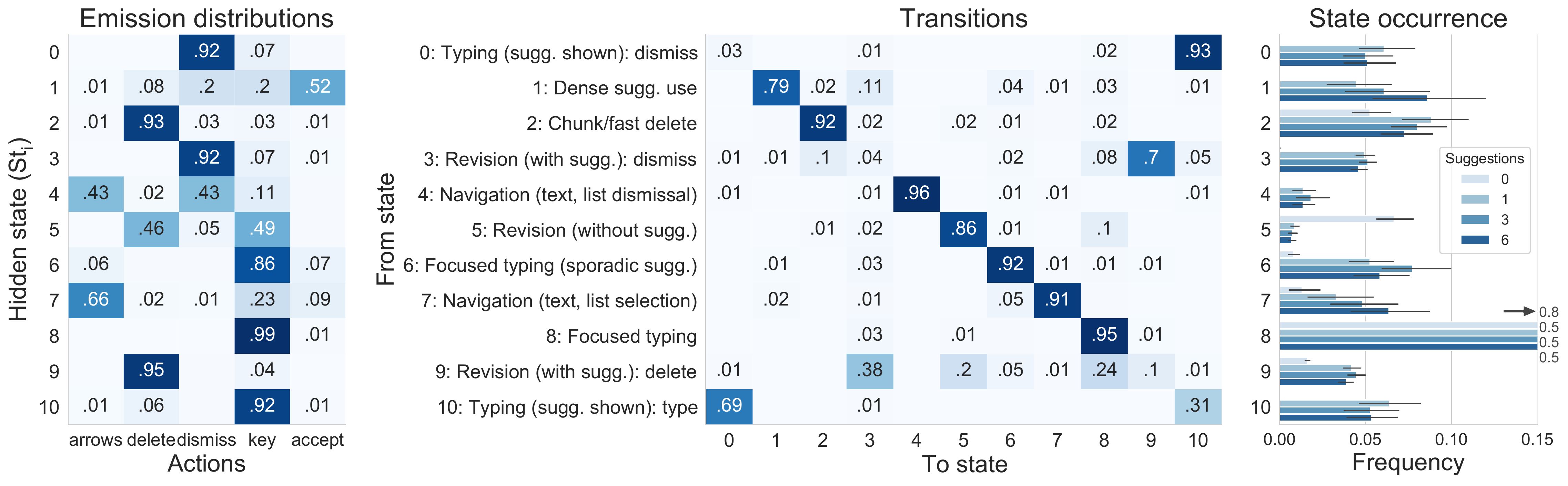}
\vspace{-2.5em}
\caption{A Hidden Markov Model (HMM) representing sequential behaviour in our study: Emission distributions of five actions (left), and state transition probabilities (centre), fitted on the data. State labels were derived by interpreting these parameters (cf. Table~\protect\ref{tab:hmm_interpretation}). The right plot shows state frequencies in the dataset, as decoded by the HMM with the Viterbi algorithm~\protect\cite{Barber2012}.}
\label{fig:hmm}
\Description[Visualisation of Hidden Markov Model parameters and decodings on the dataset.]{Left plot: Emission distributions of five actions (matrix plot / heatmap). Centre plot: State transition probabilities (matrix plot / heatmap), fitted on the data. State labels were derived by interpreting these parameters. Right: Bar plot showing state frequencies in the dataset, as decoded by the HMM with the Viterbi algorithm.}
\end{figure*}

\begin{table*}[!t]
\centering
\scriptsize
\renewcommand{\arraystretch}{1.75}
\setlength{\tabcolsep}{3pt}
\begin{tabularx}{\textwidth}{>{\raggedright}p{4.5em} >{\raggedright}p{7em}p{2.5em}>{\raggedright}p{30em}X}
    \toprule
    & \textbf{Behaviour} & \textbf{States} & \textbf{Behaviour description / interpretation of involved state(s)} & \textbf{Explanation of this interpretation, based on HMM parameters (cf. Figure~\protect\ref{fig:hmm})} \\ 
    \midrule

    Typing & Focused typing &  
    \hmmstate{8} &
    The behaviour of entering text manually without waiting for suggestions to be displayed. Also: Typing without suggestions enabled. &    
    Action \textit{\actionKey} highly likely in \hmmstate{8}, likely many in a row (high \hmmstate{8}-to-\hmmstate{8} transition). \\
    
    & Typing\linebreak(sugg. shown) &  
    \hmmstate{0}, \hmmstate{10}  &
    Entering text manually and viewing suggestions. Also: Typing focused yet slow enough for suggestions to appear. &    
    \hmmstate{0} and \hmmstate{10} form a cycle (high \hmmstate{0}-to-\hmmstate{10} and back), alternating \textit{\actionKey} and \textit{\actionDismiss}. \hmmstate{10}-to-\hmmstate{10} is also high, i.e. often multiple key presses between dismissals of suggestions. \\
    \midrule
    
    Sugg. use & Sporadic use &  
    \hmmstate{6} &
    Similar to focused typing (\hmmstate{8}): Entering text manually -- yet here sometimes accepting a suggestion. &    
    Action \textit{\actionKey} most likely, yet \textit{\actionArrow} and \textit{\actionSelect} also considerably likely (6-\pct{7}); likely multiple such actions in a row (high \hmmstate{6}-to-\hmmstate{6}).  \\
    
    & Dense sugg. use &  
    \hmmstate{1} &
    Accepting multiple suggestions in a row or with few keys in between. &    
    Action \textit{\actionSelect} highly likely for \hmmstate{1}, and likely multiple such actions (high \hmmstate{1}-to-\hmmstate{1}). \\
    \midrule

    Navigation & List selection &  
    \hmmstate{7} &
    The behaviour of navigating through a suggestion list. &    
    Likely actions \textit{\actionKey}, \textit{\actionArrow}, \textit{\actionSelect}, i.e. typing, navigating list, selecting suggestions.\\ %
    
    & Text navigation &
    \hmmstate{4}, \hmmstate{7} &
    Navigating through the text with the arrow keys. &
    Most likely actions \textit{\actionArrow} (\hmmstate{4}, \hmmstate{7}) and \textit{\actionDismiss} (\hmmstate{4}). Note: Moving the caret left/right dismisses the current suggestions. \\
    \midrule
    
    Revision & Chunk delete & 
    \hmmstate{2} &
    Repeated backspacing/deletion to remove a (larger) piece of text. &
    Action \textit{\actionBackspace} most likely, and likely repeated (high \hmmstate{2}-to-\hmmstate{2}). \\
    
    & Revision (without sugg.) &  
    \hmmstate{5} &
    Revising text via backspace/delete and (re-)typing, without dismissing suggestions (i.e. revision faster than sugg. delay or sugg. disabled). &    
    Actions \textit{\actionBackspace} and \textit{\actionKey} highly likely, and likely repeated (high \hmmstate{5}-to-\hmmstate{5}).  \\
    
    & Revision\linebreak(with sugg.) &  
    \hmmstate{3}, \hmmstate{9}  &
    Backspace/delete and considering suggestions. Also: Backspace/delete focused yet slow enough for suggestions to appear. &    
    \hmmstate{3} and \hmmstate{9} form a cycle (high \hmmstate{3}-to-\hmmstate{9} and back), alternating \textit{\actionBackspace} and \textit{\actionDismiss}. Several other transitions are also likely, i.e. leads over to other behaviours. \\
    
    \bottomrule
\end{tabularx}
\caption{Overview of nine behaviour patterns discovered via our HMM analysis. Columns list the involved HMM states (cf. Figure~\protect\ref{fig:hmm}), describe the behaviour, and outline how this interpretation was derived from the state parameters.}
\label{tab:hmm_interpretation}
\Description[Table listing nine behaviour patterns derived by interpreting the HMM parameters.]{Table listing nine behaviour patterns derived by interpreting the HMM parameters. Patter names are listed in the second column, textual interpretations in the fourth column.}
\end{table*}

Beyond timing, we analysed the (keyboard) data as sequences of five core actions/events: %
\textit{\actionKey} (entering text), \textit{\actionArrow} (moving the caret or moving through suggestions in the list with arrow keys), \textit{\actionBackspace} (backspace/delete), \textit{\actionDismiss} (dismissing suggestion list, e.g. by typing), \textit{\actionSelect} (accepting a suggestion).
These larger categories allowed us to look for fundamental patterns in the order/sequences of actions. To further facilitate this, we used a Hidden Markov Model (HMM) as a sequence analysis tool, loosely inspired by process mining and network analysis in writing research (e.g.~\cite{Leijten2013}).

\subsubsection{HMM Analysis}
An HMM~\cite{Barber2012, Rabiner1989} assumes that a sequence of $T$ observations $o_{1:T}$ (here: actions) results from a sequence of latent (``hidden'') states $h_{1:T}$ (here: higher-level behaviours that we intend to discover). Each state emits each observation with probability $p(o|h)$ and state transitions happen with $p(h_t|h_{t-1})$ (Markov chain). 
The number of hidden states is a hyperparameter, which we informed on a small set of values, using half of the people's data for fitting, the rest for evaluation (11 states scored best, considering log-likelihood and BIC~\cite{Celeux2008}).
We then fitted an HMM\footnote{We used \textit{hmmlearn} (0.2.3): \url{https://hmmlearn.readthedocs.io/}, \lastaccessed} on all action sequences, to analyse its states as more fundamental patterns of behaviour, compared to looking at raw key presses. %

\subsubsection{Intepreting Behaviour Patterns}
The fitted HMM can be seen as a summary of sequential behaviour in our study (\cref{fig:hmm}) and indicated nine patterns, as described in Table~\ref{tab:hmm_interpretation}. 
We checked these interpretations by looking at sequences in which the HMM inferred such states to occur (i.e. using the Viterbi algorithm to find the most likely hidden state sequences~\cite{Barber2012}). To check for local minima, we also compared a few repeated HMM fits with random initialisations. We observerd the behaviours to be similarly present in the learned states across these fits. 
Nevertheless, the behaviours emerging here should be seen as an \textit{exploratory} result, not a confirmatory one.

\subsubsection{Summary of Patterns}
The behaviours in \cref{tab:hmm_interpretation} relate to typing, suggestion use, navigation, and revision.
Regarding dismissals, note that not every dismissal means that the user attended to the list (e.g. list dismissed because user continues typing without looking at it).
The key finding here is that the patterns reveal \textit{varying engagement} with suggestions: 
We found focused manual typing where suggestions are mostly ignored or not triggered at all. For illustration, this might be a burst of typing following a new thought. 
On the other end, we found sporadic and heavy use of suggestions, including multiple ones in a row. These patterns indicate both user-driven and suggestion-driven ways of composing. 
A detailed look at the sequences showed that variations occur both between users and within users over the course of writing.

\begin{figure*}[!t]
\centering
\includegraphics[width=\linewidth]{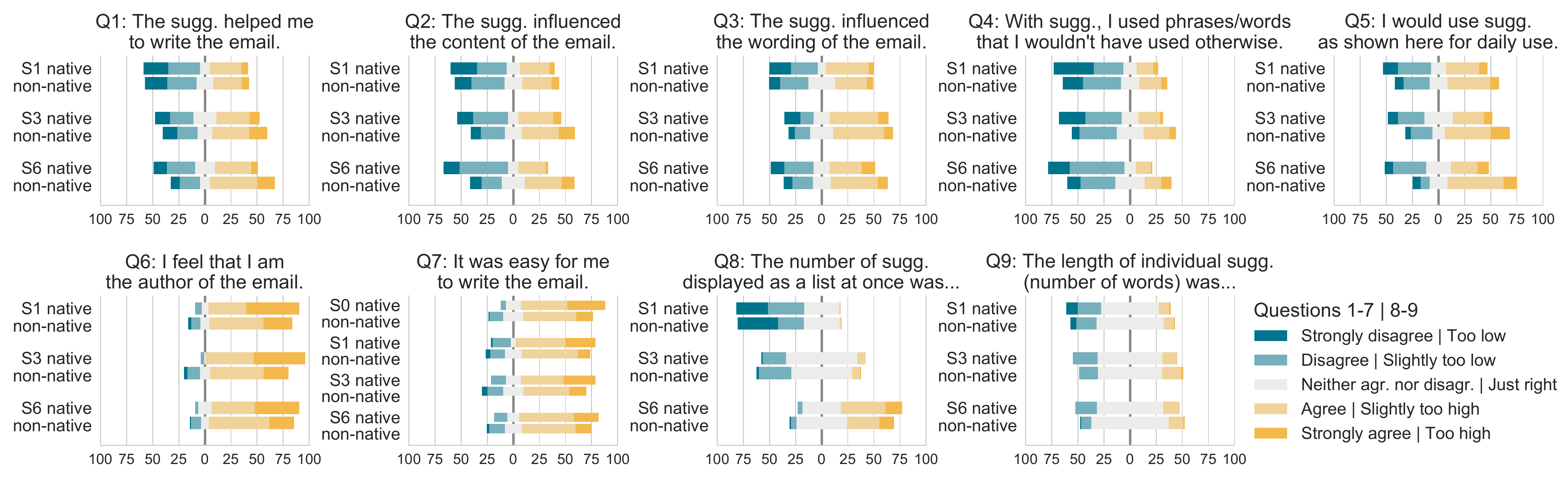}
\vspace{-2.5em}
\caption{Overview of Likert results from the questionnaires after each task. Per question, groups of bars show ratings per number of \ivsuggestions, with one bar per level of \ivlanguage (native vs non-native English speaker). The x-axes show respondents in percent. Q1-7 use agreement, Q8-9 use ``too low'' to ``too high'' (see legend).}
\label{fig:likert_agreements}
\Description[Diverging bar charts showing Likert results from the questionnaires after each task.]{Diverging bar charts showing Likert results from the questionnaires after each task. Per question, groups of bars show ratings per number of parallel suggestions, with one bar per level of language proficiency (native vs non-native English speaker). The x-axes show respondents in percent. Q1-7 use an agreement scale, Q8-9 use ``too low'' to ``too high''. Key numbers are in the text.}
\end{figure*}

\subsection{Suggestions and Text Content}

We ranked word 4-grams per scenario by occurrence ratio when writing without suggestions vs with them. %
For \textit{Birthday}, for example, 4-grams with ``lovely'' (e.g. ``have a lovely day'') were 14 times as common without suggestions as with them -- while 4-grams from ``hope you are doing well'' were 4-12 times as common with suggestions.
For \textit{Interview}, 4-grams referencing prior contact (e.g. ``we just spoke on'', ``Following on from our'') were 10-12 times as common without suggestions -- while 4-grams from ``let me know if you have any questions'' were 3-10 times as common with suggestions.
We also found examples for the other scenarios (e.g. ``please let me know'' 10-20 times more common with suggestions in both \textit{ID Card} and \textit{Reference}). 
While clearly study-specific, these examples illustrate that suggestions might replace other wordings (cf.~\cite{Arnold2020}). %

\subsection{Perception and Subjective Feedback}

Per task, we asked about perception using Likert items (\cref{fig:likert_agreements}).

\subsubsection{Perception of Self, Task and Suggestions (Q1-7)}\label{sec:results_likert_q1to7}

Descriptively (\cref{fig:likert_agreements}), many questions received mixed responses, suggesting individual differences in use, perception and preferences. Due to a technical issue the two questions added after the prestudy were not displayed for the first 35 people (i.e. N=121 for Q2, Q3). The GEE analysis found \ivlanguage{} (but not \ivsuggestions) as a significant predictor for three of the questions:

For Q4, the odds of a higher rating by non-native speakers were 2.04 times those of native speakers (p<.05): Non-native speakers were significantly more likely to perceive to a larger extent that they used other phrases/words due to suggestions. However, overall tendency was disagreement (median ``neutral''/3 for non-natives, ``disagree''/2 for natives).

For Q6, the odds of a higher rating by non-native speakers were 0.42 times the odds of native speakers (p<.01): Non-native speakers were significantly more likely to perceive themselves as authors less when writing with suggestions. To examine this further, we computed per person 1) their median rating of perceived authorship and 2) the person's mean number of suggestion acceptances. The correlation was significant (Kendall's $\tau$=-.28, p<.0001), indicating that people who accepted more suggestions also rated perceived authorship as lower. However, in absolute ratings, there was agreement with feeling to be the author (median ``agree''/4 for both).

For Q7, the odds of a higher rating by non-native speakers were 0.40 times the odds of native speakers (p<.005): Native speakers were significantly more likely to perceive writing as easier, yet absolutely, both found it rather easy (median ``agree''/4 for both).

\subsubsection{Opinion on Suggestion Settings (Q8-9)}
In Q8, \pct{62} of people rated three suggestions as ``just right''/3 (median). For one/six suggestions, more people found them too few/many (median ''too low''/2 and ''too high''/4), respectively, yet also many rated ``just right'' (\pct{34} for $\suggone$, \pct{44} for $\suggsix$).
The number of words per suggestion (Q9) was perceived as ''just right''/3 (all medians), with a slight indication that a single suggestion could be longer for some users.

\subsubsection{Open Feedback}
We asked what people liked or disliked. %
One author inductively developed a codebook and coded all responses. Another author iterated on the codebook and also coded all responses. We resolved differences in these codings via discussion. 

Most frequent positive aspects were overall good suggestions (\pct{38.5} of people), being helpful (\pct{24.4}) and providing inspiration (\pct{22.4}). Also mentioned were usability (\pct{20.5}), efficiency (\pct{15.4}), suggestion variety (\pct{10.9}), and help with wording (\pct{9.0}) and when feeling stuck (\pct{9.0}). Nine people (\pct{5.8}) stated nothing positive.

Top mentioned aspects for improvement were suggestion quality (\pct{25}; e.g. relevance or grammar) and better considering email context and topic (\pct{12.8}). Twelve people (\pct{7.7}) would have liked shorter suggestion delay or faster loading, and eight (\pct{5.1}) mentioned distraction. People also mentioned more controllable parameters (\pct{8.3}) and related ideas, such as a customisable suggestion vocabulary (\pct{5.1}) or selecting topic, mood or level of formality. About a third stated that there is nothing to improve (\pct{36.5}).

\section{Discussion}

\subsection{Multiple Suggestions Help to Find Phrases}
Multiple suggestions support finding useful phrases, as indicated by several results: 
More parallel suggestions increased their acceptance rate (even more so for non-native speakers). More suggestions also decreased the need for their manual modification  (\cref{fig:metrics_overview}). 
Choice (i.e. >1 suggestion) also led to more suggestion words in the email, without increasing email length. %
These results suggest that people get value out of seeing a set of phrases, instead of one, and that they selectively look for ``right'' phrases. 
Finally, subjective feedback (\cref{fig:likert_agreements} Q8) showed high acceptance also in tasks with three (\pct{62} found it ``just right'') and six (\pct{44} found it ``just right'') suggestions. 

Thus, our results here motivate to not always rely on the currently dominant default of suggesting one phrase or three words (\cref{sec:ui_design_factors}), but rather explore a larger range, at least for design goals beyond speed and when addressing specific user groups. %

\subsection{User Engagement with Suggestions Varies}

People use suggestions selectively and with varying engagement: 
Behaviour patterns indicate a wide range (\cref{tab:hmm_interpretation}), from typing without suggestions, over sporadic integration, to chaining multiple ones. Thus, people seem to vary between manual and suggestion-driven composition.
They further edited \pct{21.3} of accepted suggestions -- another behaviour with rather high engagement with suggestions.
Future studies could examine if engagement varies due to experiences with the system, its quality, or user strategies.

The nine behaviour patterns emerging here (\cref{tab:hmm_interpretation}) are the first such set described in the literature in this context, and thus provide a starting point for replication and exploration. Beyond suggestions, they motivate supporting \textit{revision}. This highlights the relevance of recent work on (mobile) text revision~\cite{Komninos2018, Li2020, Zhang2019} and of calls for research into text ``interaction'' instead of ``entry''~\cite{Vertanen2017}.

\subsection{Suggestions Cost Time and Actions}
People tended to take longer with three and six suggestions, particularly native speakers (\cref{fig:metrics_overview}).
Similarly, recent work on smartphones~\cite{Palin2019} found word prediction to negatively correlate with speed, yet noted that the variability between individuals motivates a more detailed look at users. We provide such detail -- revealing language proficiency as one relevant factor in this context.

We also provide a first estimate of time costs of (four-word) phrase choice: +\ms{610} per suggestion beyond the first ($R^2$=.98; fitted on the mean search \& selection times in \cref{fig:metrics_overview}). In menu search~\cite{Bailly2014}, menu length also influences search time. This is logarithmic according to the Hick-Hyman law~\cite{Hyman1953, Landauer1985}. However, menus with fixed items benefit from familiarisation and recognition, while suggestions change, likely making it more difficult to go beyond serial search. Still, future work could explore if users become familiar with a language model (e.g. develop useful expectations). %

Suggestions also incur action costs. %
In our analysis, focused typing accounts for \pct{80} of (keyboard) actions/events if suggestions are disabled (other actions are e.g. revision). This goes down to about \pct{50} with suggestions (cf. \cref{fig:hmm} right), as some actions shift to navigating/selecting suggestions or reading/dismissing them. Note that we do not know if every list dismissal means that the user indeed attended to the list; %
future work could use eye tracking to assess this. 
Nevertheless, these results show that behaviour patterns are considerably impacted by suggestions in the UI.

Related work addressed the costs of attending to suggestions with utility-gated suggestions~\cite{Quinn2016chi} or by showing only one suggestion~\cite{Chen2019}. 
Our results support such approaches if the main goal is to save time, particularly when assuming native speakers. However, %
suggestions are also valued beyond efficiency, as discussed next.

\subsection{Users Consider More than Efficiency}\label{sec:beyond_efficiency}
People in our study indeed valued suggestions beyond saving keystrokes or time, in line with our expectations based on related work (also see intro):
First, despite an implicit incentive for speed in the study (fixed pay), people chose to invest several seconds into choosing suggestions. Still, \pct{15.4} in open comments mentioned (perceived) efficiency benefits.  %
Second, \pct{22.4} positively commented on aspects of inspiration, and \pct{9.0} said they found help when stuck. Many also perceived suggestions to influence wording and content (\pct{49.3} and \pct{41.3}, respectively, cf. Figure~\ref{fig:likert_agreements}), and open comments indicated that this influence was seen as positive.
Feedback and perception in the prestudy also support this use for ideation.
Differences between native and non-native speakers (cf. Figure~\ref{fig:metrics_overview}) further point at a supporting role beyond a purely execution-related one. 

Such analyses beyond efficiency are still sparse, as also evident from workshop calls motivating work beyond speed (e.g.~\cite{Vertanen2016}).
Related work also found aspects of ideation, comparing word and phrase suggestions for mobile text entry~\cite{Arnold2016uist}.
Interestingly, related observations also appear for code suggestions (in IDEs), which are valued not only to reduce typing yet also to learn and explore an API~\cite{Omar2012, Robbes2008}.
Adding to the literature, our study thus for the first time provides evidence of use and perception of (multiple) phrase suggestions beyond efficiency, for emailing, at the desktop.%

\subsection{Language Proficiency Matters}

Our study revealed clear differences in suggestion use depending on language proficiency: Non-native speakers accepted and used more suggestions and based on these metrics gained relatively more from seeing more parallel suggestions (cf. acceptance rates in Figure~\ref{fig:metrics_overview}). Related, temporal costs of suggestions seemed to be less of an overhead for non-native speakers (cf. task times in \cref{fig:metrics_overview}). Non-native speakers were also overall slightly more positive about the helpfulness of suggestions and also perceived their influence on wording and content, and inspiration for using other phrases/words, as (slightly) higher (cf. \cref{fig:likert_agreements} and \cref{sec:results_likert_q1to7}). %

These findings motivate %
further research into non-native predictive text entry, for instance, to investigate why non-native speakers accept more suggestions (e.g. more helpful vs more difficult to spot problems). This aligns well with calls for text entry research to consider varied user groups and go beyond speed~\cite{Clawson2015, Kristensson2013, Vertanen2016, Vertanen2017}. Moreover, although aspects of it do appear~\cite{Clawson2015}, language proficiency often seems underrepresented (e.g. compared to mentionings of impairments or age~\cite{Vertanen2017}). In this context, our study provides a first motivating comparison and dataset of native and non-native use of phrase suggestions, also to stimulate further work.

\subsection{Implications for Research and Design}\label{sec:implications_for_uis}

\subsubsection{Designing for Choice}
As the study showed, parallel suggestions can have value yet cost time/action. 
Future designs could aim to support \textit{comparing} suggestions at a glance (e.g. optimise order, highlight keywords). For long suggestions, suggested text might even be summarised for faster parsing. Designs could also go beyond lists (e.g. hierarchical menu with semantic categories).

\subsubsection{Navigating Suggestions}
Suggestion lists need to be navigated (our prototype: arrow keys + enter, or tab to accept first suggestion). While we also offered mouse selection, this may lead to costly keyboard-mouse switches, also mentioned in the comments. Future work could study further interactions (e.g. direct selection via numbers 1-X, filtering by continued typing, gaze-based selection).

\subsubsection{Triggering Suggestions}
We showed suggestions with a slight delay after a keypress. This was accepted by participants, yet our results also motivate design explorations. 
For example, some found this distracting, wished for shorter delays or for an adaptive approach. This might be addressed with utility-based suggestion triggers (e.g.~\cite{Chen2019, Quinn2016chi}) or adapting to input behaviour and context, as discussed next. Comparison to an explicit trigger (e.g. ``tab`` to get suggestions) might also be insightful.

\subsubsection{Adaptive and Adaptable UIs}
Our results motivate adaptive UIs (e.g. on/off, number of suggestions, length), for example, to account for changing user-driven or suggestion-driven composition strategies. Adaptation could be based on current behaviour (e.g. deactivate during focused typing), possibly inferred from input (see our HMM analysis) and further sensors (e.g. eye-tracking). Related, the results and people's feedback also motivate user-controlled adaptation, including ideas for selecting a topic, mood or level of formality for suggestions. 
Given the differences depending on language proficiency, a single user might also benefit from changeable designs or settings if they use multiple languages. %

\subsubsection{Generating Suggestions as a Set}
Our results motivate further work on generating phrases with a focus on their variety and utility \textit{as a set} (cf.~\cite{Arnold2016uist, Deb2019, Kannan2016}), instead of focusing on (single) most likely phrases or next words. Further supporting this idea, a few people explicitly wished for more variety in meaning, instead of wording.

\subsection{Limitations \& Reflections on  Methodology}

Study behaviour might differ from real life. To mitigate this, we chose typical (business) cases -- and the duration and emails showed that people invested overall realistic time and effort. 
We also used a composition task, for which there is evidence that it has better external validity than text copy tasks, also as an online study~\cite{Vertanen2014}.
There may also be novelty effects: While \pct{85} reported to use suggestions at least sometimes, multiple phrase suggestions were likely new to many. Related, we elicited four emails per person and already found that people got faster, both in writing (1st task $M$=\secs{260}, 4th task $M$=\secs{181}) and in selecting suggestions (1st task $M$=\ms{4309}, 4th task $M$=\ms{3656}).
While the study amounted to ca. 20 minutes of use already, we motivate future studies to compare our results to more long-term (and ``in the wild'') observations.

Our model is a prototype. Suggestion quality could be further improved, for instance, through further finetuning or training efforts, possibly involving even larger (email) datasets, extensive architecture search, or generally scaling up (e.g.~\cite{Brown2020language}). Performance might be further improved with further serving and scaling efforts.

Moreover, we found interesting differences in a comparison of native vs non-native language proficiency, %
which motivates future study of more varied language levels.
Related, our data is made available and affords further analyses, for instance, on linguistic interests (e.g. fluency indicators, such as pauses/bursts, cf.~\cite{Leijten2019}). 

Finally, the rich dataset and insights gained here can be seen as evidence for text composition as a useful study task~\cite{Vertanen2014}, also considering calls for work that goes beyond text copy tasks (e.g.~\cite{Vertanen2016}).

\subsection{Broader Reflections on Text Suggestions}

\subsubsection{Quality}

Despite prestudy and finetuning efforts, some suggestions were unsuitable. Some people mentioned grammar and topic/formality as areas for improvement. This is an active research area: For example, \textit{GPT-3}~\cite{Brown2020language} was published while writing this paper. A future study could compare its use to our results here. %
For communication, quality perceived by the receiver could also be studied. Related, emailing is more goal-directed than, for example, chat messaging, which might influence needs and preferences.

\subsubsection{Bias}
Quality, beyond the technical~\cite{Blodgett2020}, also involves undesirable or biased suggestions (e.g. swear words, gender stereotypes). %
Even building on a published model, and in a business context, at least two people saw inappropriate words. This motivates further work, %
especially where end-users face model output. Related efforts also ask for further model documentations~\cite{Mitchell2019}.

For \textit{interactive} use of language models, our results indicate that bias may not be limited to the model but may also manifest in UI design: Concretely, a design focus on efficiency (e.g. single suggestion, gated by estimated keystroke/time savings) may obstruct other supporting effects for users with lower language proficiency.

\subsubsection{Learning vs Deskilling}
Our study gives a glimpse at two potential impacts on writing skills: 
On the one hand, data and feedback indicate that suggestions may help with phrasing, wording and spelling, in particular in a foreign language. Long-term studies could investigate if this improves proficiency, even when writing \textit{without} suggestions, as there is evidence for learning with spell checking and autocorrection~\cite{Arif2016, Fleming2019}.
On the other hand, reliance on suggestions could also have a negative impact on writing skills.%

Related, while people in both prestudy and main study referred to text influence as positive, it might also %
limit creativity: For instance, a recent study found text suggestions to make image captions more predictable~\cite{Arnold2020}. %
Such tensions are not new, nor limited to text suggestions~\cite{Carr2015}. In a broad view, we see these considerations here as further motivation for developing interactive AI from an \textit{interdisciplinary} perspective that considers broad consequences.%

\section{Conclusion}

As our key empirical contribution to the literature, our study for the first time compared writing emails with different numbers of parallel phrase suggestions, and for native and non-native English writers. To conclude, put simply, one might thus ask: Are multiple phrase suggestions worth it, for what, and for whom?

Based on our study, we overall conclude that suggesting multiple phrases is useful for ideation (i.e. coming up with what to write next and how to write it), at a cost of efficiency. While we observed this for both native and non-native speakers, the latter overall seemed to benefit more from more parallel suggestions.

In a broad view, our results challenge research in (predictive) text entry to look  beyond previous focus areas, in line with calls from the community (e.g.~\cite{Clawson2015, Vertanen2017}). Our study empirically responds to such calls and our findings particularly motivate: 
(1) Exploring a larger variety of system/UI parameters (e.g. beyond suggesting one phrase or three single words). 
(2) Exploring design goals beyond efficiency, such as ideation or language learning. 
(3) Explicitly considering skills, needs and preferences of specific user groups.

Moreover, we found nine fundamental behaviour patterns related to text input, navigation, revision and suggestion use. Beyond our focus on suggestions here, these motivate work on supporting text interactions beyond text entry, in particular revision.

Finally, by motivating further design considerations here, we hope to contribute to future AI systems that support and augment human writers, instead of replacing them.
To facilitate further research in this area, we provide the prototype, model, data and study material on the project website:
\url{https://osf.io/7q4c8/}

\begin{acks}
We thank Fiona Draxler and Mark Dunlop for feedback on the manuscript, and Robin Welsch for feedback and statistical consulting. %
This project is funded by the Bavarian State Ministry of Science and the Arts and coordinated by the Bavarian Research Institute for Digital Transformation (bidt).
\end{acks}

\bibliographystyle{ACM-Reference-Format}
\bibliography{bibliography}

\end{document}